\documentstyle[12pt,epsfig]{article}
\begin{document}
\centerline{\Large\bf Chaotic inflation with a quadratic potential
in all dimensions}
\vspace*{0.050truein}
\centerline{Forough
Nasseri\footnote{Email: nasseri@fastmail.fm}}
\centerline{\it Department of Physics, Neishabour University,
P.O.Box 769, Neishabour, Iran}
\centerline{\it Khayyam Planetarium, P.O.Box 769, Neishabour, Iran}
\begin{center}
(\today)
\end{center}
\begin{abstract}
We study chaotic inflation with a quadratic potential
in all dimensions. The slow-roll parameters, the spectral indices
of scalar and tensor
perturbations and also their running have been calculated
in all dimensions.
\end{abstract}

Following the advent of string theory and its implication that space
may have more than the usual three dimensions, we here study chaotic
inflation with a quadratic potential, $V(\varphi)=m^2 \varphi^2/2$,
in all dimensions.

Take the metric in constant $(D+1)$-dimensional spacetime in the
following form (we use natural units or high-energy physics units in which
the fundamental constants are $\hbar=c=k_B=1$, $G=\ell_{Pl}^2=1/m_{Pl}^2$)
\begin{equation}
\label{1}
ds^2=-N^2(t) dt^2 + a^2(t) d \Sigma_k^2,
\end{equation}
where $N(t)$ denotes the lapse function and $d \Sigma_k^2$ is the line
element for a $D$-manifold of constant curvature $k=+1,0,-1$,
corresponding to the closed, flat and hyperbolic spacelike sections,
respectively. The Ricci scalar is given by \cite{1}
\begin{equation}
\label{2}
R= \frac{D}{N^2} \left\{ \frac{2{\ddot a}}{a} + (D-1)
\bigg[ \left( \frac{\dot a}{a} \right)^2 + \frac{N^2 k}{a^2} \bigg]
- \frac{2 {\dot a} {\dot N}}{a N} \right\}.
\end{equation}
The Einstein-Hilbert action for the pure gravity is given by \cite{1}
\begin{equation}
\label{3}
S_G= \frac{1}{2 \kappa_{(D+1)}} \int d^{(D+1)} x \sqrt{-g} R,
\end{equation}
and the action for a perfect fluid may be expressed by
\begin{equation}
\label{4}
S_M= - \int d^{(D+1)} x \sqrt{-g} \rho.
\end{equation}
In Eq.(\ref{3}), the gravitational coupling constant in $(D+1)$-dimensional
spacetime, $\kappa_{(D+1)}$, is related to the $(D+1)$-dimensional
gravitational constant $G_{(D+1)}$ by \cite{2}
\begin{equation}
\label{5}
\kappa_{(D+1)}=(D-1) S^{[D]} G_{(D+1)},
\end{equation}
where
\begin{equation}
\label{6}
S^{[D]}= \frac{2 \pi^{D/2}}{\Gamma \left( \frac{D}{2} \right)},
\end{equation}
where $S^{[D]}$ is the surface area of the unit sphere in $D$-dimensional
spaces. In the case $(3+1)$, $(4+1)$ and $(5+1)$-dimensional spacetime we
have $\kappa_{(3+1)}=8 \pi G_{(3+1)}$ (i.e. $\kappa= 8 \pi G$),
$\kappa_{(4+1)}= 6 \pi^2 G_{(4+1)}$ and
$\kappa_{(5+1)}=\frac{32 \pi^2}{3}G_{(5+1)}$, respectively.
Using (\ref{5}) and (\ref{6})
we have (See Appendix)
\begin{equation}
\label{7}
\kappa_{(D+1)}=
\frac{2 (D-1) \pi^{D/2} G_{(D+1)}}{\Gamma \left( \frac{D}{2} \right)}.
\end{equation}
Using (\ref{3}) and (\ref{4}), we obtain the following Lagrangian \cite{1}
\begin{equation}
\label{8}
L:= - \frac{a^D}{2 \kappa_{(D+1)}} \frac{D(D-1)}{N}
\left[ \left( \frac{\dot a}{a} \right)^2 - \frac{N^2 k}{a^2} \right]
-\rho N a^D.
\end{equation}
Varying the above Lagrangian with respect to $N$ and $a$, we find the
following equations of motion in the gauge $N=1$, respectively
\begin{eqnarray}
\label{9}
&&\left( \frac{\dot a}{a} \right)^2 + \frac{k}{a^2}
=\frac{2 \kappa_{(D+1)} \rho}{D(D-1)},\\
\label{10}
&&\frac{\ddot a}{a} + \left[ \left( \frac{\dot a}{a} \right)^2
+ \frac{k}{a^2} \right] \left( -1 + \frac{D}{2} \right) +
\frac{\kappa_{(D+1)} p}{D-1} =0.
\end{eqnarray}
Using (\ref{9}) and (\ref{10}), one gets the continuity equation
\begin{equation}
\label{11}
\frac{d}{dt} \left( \rho a^D \right) + p \frac{d}{dt} \left( a^D \right)
=0.
\end{equation}
Inserting $k=0$ for a flat universe and the energy density and the
pressure of a homogeneous inflaton field
\begin{eqnarray}
\label{12}
\rho &\equiv& \frac{1}{2} {\dot \varphi}^2 + V(\varphi),\\
\label{13}
p &\equiv& \frac{1}{2} {\dot \varphi}^2 - V(\varphi),
\end{eqnarray}
in Eqs.(\ref{9}) and (\ref{11}), we are led to
\begin{eqnarray}
\label{14}
&&H^2 
=\frac{2 \kappa_{(D+1)}}{D(D-1)}
\left( \frac{1}{2} {\dot \varphi}^2 + V(\varphi) \right),\\
\label{15}
&&{\ddot \varphi} + D H {\dot \varphi} = -V'(\varphi),
\end{eqnarray}
where $H=\frac{\dot a}{a}$. The slow-roll conditions in
$D$-dimensional spaces read \cite{3}
\begin{equation}
\label{16}
{\dot \varphi}^2 \ll V (\varphi),\;\;\;\;
{\ddot \varphi} \ll D H {\dot \varphi},\;\;\;\;
-{\dot H} \ll H^2.
\end{equation}
Using these conditions, Eqs.(\ref{14}) and (\ref{15}) can be rewritten
\begin{eqnarray}
\label{17}
H^2 =
\frac{2 \kappa_{(D+1)} V(\varphi)}{D(D-1)},\\
\label{18}
D H {\dot \varphi} \simeq - V'(\varphi).
\end{eqnarray}
During inflation, $H$ is slowly varying in the sence that its change per
Hubble time, $\epsilon \equiv -{\dot H}/H^2$ is less than one. The
slow-roll condition $\eta\ \ll 1$ is actually a consequence of the
condition $\epsilon \ll 1$ plus the slow-roll approximation
$D H {\dot \varphi} \simeq - V'(\varphi)$. Indeed, differentiating
(\ref{18}) one finds
\begin{equation}
\label{19}
\frac{\ddot \varphi}{H {\dot \varphi}} = \epsilon - \eta,
\end{equation}
where the slow-roll parameters in any constant space dimension are defined
by
\begin{eqnarray}
\label{20}
\epsilon & \equiv & - \frac{\dot H}{H^2} = \frac{(D-1)}{4 \kappa_{(D+1)}}
\left( \frac{V'}{V} \right)^2,\\
\label{21}
\eta & \equiv & \frac{V''}{D H^2} = \frac{(D-1)}{2 \kappa_{(D+1)}}
\left( \frac{V''}{V} \right).
\end{eqnarray}
The number of e-foldings between $t_i$ and $t_f$ is given by
\begin{equation}
\label{22}
{\cal N} = \int_{t_i}^{t_f} H(t) dt=
\ln \left( \frac{a_f}{a_i} \right) \simeq
- \frac{2 \kappa_{(D+1)}}{(D-1)} \int_{\phi_i}^{\phi_f}
\frac{V}{V'} d \varphi.
\end{equation}
The amplitudes of scalar and tensor perturbations generated in inflation
can be expressed by \cite{{4},{5},{6}}
\begin{eqnarray}
\label{29}
A_S^2 &=& \left( \frac{H}{2 \pi} \right)^2 \left( \frac{H}{\dot \varphi} \right)^2,\\
\label{30}
A_T^2 &=& \frac{\kappa_{(D+1})}{(D-1)} \left( \frac{H}{2 \pi} \right)^2.
\end{eqnarray}
These amplitudes are equal to $\frac{25}{4}$ times the amplitudes as
given in Ref.\cite{5}. The amplitudes of scalar and tensor perturbations
generated in inflation can be determined by substituting (\ref{17})
and (\ref{18}) into (\ref{29}) and (\ref{30})\footnote{It is worth
mentioning that the authors of Refs.\cite{{3},{6}} have studied chaotic
inflation in higher dimensions and also in a model universe with time
variable space dimensions by taking $\kappa=8 \pi G$ for all dimensions.
Our results above improve the results given in Refs.\cite{{3},{6}}
because we here consider the gravitational coupling constant in all
dimensions as a function of spatial dimensions, as given in Eq.(\ref{7}).}
\begin{eqnarray}
\label{31}
A_S^2 &=& \frac{ 2 \kappa_{(D+1)}^3 V^3}{D(D-1)^3 \pi^2 V'^2},\\
\label{32}
A_T^2 &=& \frac{\kappa_{(D+1)}^2 V}{2 \pi^2 D(D-1)^2}.
\end{eqnarray}
These expressions are evaluated at the horizon crossing time when $k=aH$.
Since the value of Hubble constant does not change too much during
inflationary epoch, we can obtain $dk=H da$ and $d \ln k = H dt = da/a$.
Using the slow-roll condition in $(D+1)$-dimensional spacetime
\begin{equation}
\label{33}
\frac{d}{d \ln k}=-\frac{V'}{D H^2} \frac{d}{d \varphi},
\end{equation}
and also a lengthy but straightforward calculation by using
(\ref{17}), (\ref{18}), (\ref{20}), (\ref{21}) and
(\ref{29})-(\ref{33}) we find
\begin{eqnarray}
\label{34}
n_S-1 &\equiv& \frac{d \ln A_S^2}{d \ln k} = - 6 \epsilon + 2 \eta,\\
\label{35}
n_T &\equiv& \frac{d \ln A_T^2}{d \ln k} = - 2 \epsilon,
\end{eqnarray}
where $n_S$ and $n_T$ are the spectral indices of scalar and tensor
perturbations, respectively. If $n_S$ and $n_T$ are expressed as a
function of e-folding ${\cal {N}}$, one can use the fact that
$\frac{d}{d \ln k}=-\frac{d}{d{\cal{N}}}$ to obtain the desired
derivatives even more easily.
To calculate the running of the scalar and tensor spectral indices in
all dimensions, we use Eqs.(\ref{20}),(\ref{21}) and (\ref{33}).
Therefore we have in all dimensions 
\begin{eqnarray}
\label{36}
\frac{d \epsilon}{d \ln k} &=& - 2 \epsilon \eta + 4 \epsilon^2,\\
\label{37}
\frac{d \eta}{d \ln k} &=& 2 \epsilon \eta - \xi,
\end{eqnarray}
where the third slow-roll parameter is defined by\footnote{This expression
for $\xi$ in all dimensions improve Eq.(56) of Ref.\cite{6}
in which $\kappa=8 \pi G$ has been considered for all dimensions.}
\begin{equation}
\label{38}
\xi \equiv \frac{(D-1)^2}{4 \kappa_{(D+1)}^2} \left( \frac{V' V'''}{V^2}
\right).
\end{equation}
Using the above equations, running of the spectral indices of scalar and
tensor perturbations in higher dimensions have these explicit expressions
\begin{eqnarray}
\label{39}
\frac{d n_S}{d \ln k} &=& 16 \epsilon \eta - 24 \epsilon^2 - 2 \xi,\\
\label{40}
\frac{d n_T}{d \ln k} &=& 4 \epsilon \eta - 8 \epsilon^2.
\end{eqnarray}

For the chaotic inflation with a quadratic potential, $m^2 \varphi^2/2$,
the solution of Eqs.(\ref{17}) and (\ref{18}) are given by
\begin{eqnarray}
\label{23}
\varphi (t) &=& \varphi_i - m \sqrt{\frac{D-1}{D \kappa_{(D+1)}}} t,\\
\label{24}
a (t) &=& a_i \exp \left( \frac{\kappa_{(D+1)}}{2(D-1)} \left[
\varphi_i^2 - \varphi^2(t) \right] \right).
\end{eqnarray}
Using the slow-roll parameters
\begin{equation}
\label{25}
\epsilon = \eta = \frac{(D-1)}{\kappa_{(D+1)} \varphi^2},
\end{equation}
and the failure of the slow-roll conditions
\begin{equation}
\label{26}
\max \{ \epsilon_f ; |\eta_f| \} \simeq 1,
\end{equation}
one concludes that
\begin{equation}
\label{27}
\varphi_f = \sqrt{\frac{(D-1)}{\kappa_{(D+1)}}}.
\end{equation}
Substituting this value of $\varphi_f$ into (\ref{22}), one gets
\begin{equation}
\label{28}
{\cal N} = \frac{\kappa_{(D+1)}}{2(D-1)} \varphi_i^2 - \frac{1}{2}.
\end{equation}
One can also obtain the spectral indices of scalar and tensor
perturbations ans their running in ($D+1$)-spacetime dimension
\begin{eqnarray}
\label{45}
n_s-1 &=& - 4 \epsilon =-\frac{4(D-1)}{\kappa_{(D+1)} \varphi^2},\\
\label{46}
n_T &=& - 2 \epsilon =- \frac{2(D-1)}{\kappa_{(D+1)} \varphi^2},\\
\label{47}
\frac{d n_S}{d \ln k} &=& - 8 \epsilon^2 = -
\frac{8 (D-1)^2}{\kappa^2_{(D+1)} \varphi^4},\\
\label{48}
\frac{d n_T}{d \ln k} &=& - 4 \epsilon^2
= -\frac{4(D-1)^2}{\kappa^2_{(D+1)} \varphi^4}.
\end{eqnarray}
\noindent
{\bf{Appendix: Gravitational coupling constant in higher dimensions}}

In $(3+1)$-dimensional spacetime, the gravitational coupling constant
is given by $\kappa= 8 \pi G$. Looking for the roots of the factor of
$8 \pi$ in $\kappa$ we across the relation
\begin{equation}
\label{a1}
R_{00}= (D-2) \nabla^2_D \phi,
\end{equation}
where $\nabla_D$ is the $\nabla$ operator in $D$-dimensional space.
In $(3+1)$-dimensional spacetime, the Poisson equation is given by
\begin{equation}
\label{a2}
\nabla^2 \phi= 4 \pi G \rho.
\end{equation}
Applying Gauss law for a $D$-dimensional volume, we find the
Poisson equation for arbitrary fixed dimension
\begin{equation}
\label{a3}
\nabla^2_D \phi = S^{[D]} G_{(D+1)} \rho,
\end{equation}
where $S^{[D]}$ is the surface area of a unit sphere in $D$-dimensional
spaces, see Eq.(\ref{6}).
On the other hand we get
\begin{equation}
\label{a4}
R_{00} = \left( \frac{D-2}{D-1} \right) \kappa_{(D+1)} \rho.
\end{equation}
Using (\ref{6}), (\ref{a1}), (\ref{a3}) and (\ref{a4}), we are led to
the gravitational coupling constant in $(D+1)$-dimensional spacetime
as given in (\ref{5}) and (\ref{7}).

\noindent
{\bf Acknowledgments:} F.N. thanks Hurieh Husseinian and A.A.Nasseri
for noble helps and also thanks Amir and Shahrokh for truthful helps.

\end{document}